# A positron study of the Fermi surface of TmGa3

M. Biasini[1,2], A. Czopnik[3], A. Jura[3], G. Kontrym-Sznajd[3] and M. Monge[1,4]

[1]ENEA, via don Fiammelli, 2 40128 Bologna, Italy

[2]Istituto Nazionale di Fisica della Materia, Italy

[3]Trzebiatowski Institute of Low Temperature and Structure Research, P.O.Box 937, Wrocław, Poland

[4]Universidad Carlos III av Universidad 30, Leganes 28911, Spain (permanent address)

**Abstract.** For a single crystal of TmGa3, two-dimensional angular correlation of positron annihilation radiation spectra were measured and subjected to a deconvolution algorithm based on the Van Citter iterative method. The three-dimensional electron-positron momentum density was reconstructed both via the Cormack and a modified Fourier-transform-based method. After the 3D LCW transformation, the resulting $k$-space density was in fair agreement with band structure LMTO calculations. In particular, the use of deconvoluted data helped to reveal small details of the Fermi surface in the 6$^{th}$ band.

**Introduction.**

TmGa$_3$ crystallises in the cubic AuCu$_3$-type structure with lattice constant a=4.196 Å. At 4.26K it undergoes an antiferromagnetic transition [1] to the multiaxial 3**k** – type structure [2,3]. It belongs to the class of intermetallic rare-earth compounds in which the interaction between quadrupolar moments of a 4$f$ charge (via the conduction bands) is so strong that it leads to a quadrupolar ordering in the paramagnetic phase [4]. Strength and sign of this interaction depend strongly on the character of the conduction electrons near the Fermi energy.

The Fermi surface (FS) geometry and effective cyclotron masses of TmGa$_3$, determined by means of the de Haas-van Alphen (dHvA) experiment, were in agreement with self-consistent LMTO band structure calculations in a local spin density approximation [5]. In these calculations the 4$f^{12}$ states were not allowed to hybridise with 12 conduction electron states. This imposition was consistent with X-ray photoelectron spectroscopy [6], which showed that the energy of 4$f^{12}$ states lies 5eV below the Fermi energy.

We present results for reconstructed electron-positron (e-p) momentum densities in the extended **p** and reduced **k** zone schemes obtained from 2D ACAR experiment. We also derive a topology of the FS in its paramagnetic configuration phase.

**Results**

Three projections were collected at a temperature of ~60K, with integration directions at 0°, 22.5° and 45° from the [100] crystal axis (more details in [7]). Spectra were deconvoluted by applying an algorithm based on the van Citter iterative method [8] and subsequently 3D densities $\rho(p)$ were reconstructed by using a modified Fast Fourier Transform (FFT) [9] and Cormack's method (CM) [10]. Because the FFT results contained a high contribution from the noise, they are not presented here.

The reconstruction of 3D densities was reduced (as in the case of all tomography reconstruction techniques) to sets of reconstructions of 2D densities, performed, independently, on the planes perpendicular to $p_z \equiv [001]$. In our case (3 measured spectra) on each plane $p_z$=const., the 2D density

is described by:

$$\rho(p,\varphi) = \rho_0(p) + \rho_4(p)\cos(4\varphi) + \rho_8(p)\cos(8\varphi), \qquad (1)$$

where $p=(p_x^2+p_y^2)^{1/2}$ and angle $\varphi$ are defined in the polar system (for more details see ref. [9,10]). Absolute values of $\rho_4$ (in our case equal to $0.5\{\rho_{[100]}(p,\varphi=0^0)-\rho_{[110]}(p,\varphi=45^0)\}$) and $\rho_8$, compared with absolute values of $\Delta\rho_{[100]} = 0.5\{\rho_{[100]}(p,\varphi=0^0)-\rho_{[001]}(p,\varphi=0^0)\}$ are displayed in Fig. 1. Almost negligible values of $\rho_8(p)$ prove that for TmGa$_3$ three projections were quite sufficient to reproduce all details of $\rho(p)$.

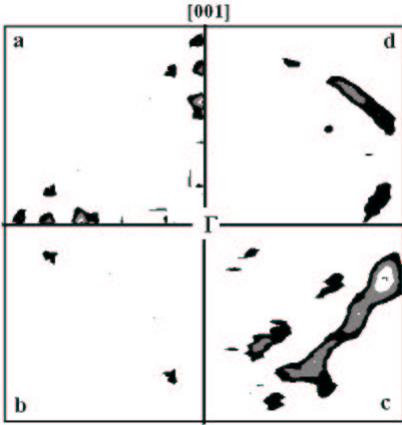

Fig. 1
Absolute values of: $\Delta\rho_{[100]}$ for standard (a) and modified (b) CM, $\rho_4(p)$ (c) and $\rho_8(p)$ (d), on the planes perpendicular to the [001] axis. We present only values greater than 0.015 (in units where maximum value of $\rho(p)$ in the central $\Gamma$ point is equal to 1) and for momenta p up to 7.2 mrad.

If there would be no experimental and reconstruction errors, $\Delta\rho_{[100]}$ should be equal to zero. Outside of the region near the [001] and [100] axes $\Delta\rho_{[100]}$ is lower than 0.02, showing mostly experimental noise (reduced by expanding data into Chebyshev polynomial series). Higher values (up to 0.06) along (and near) the main symmetry axes are connected with the fact that reconstructed densities for small values of $p$ contain the highest "reconstruction errors". Since values of $\rho_4(p)$ are of the same order but these small quantities have an essential influence on $\rho(k)$ (densities folded into the first BZ [11]), we modified CM reducing the experimental and reconstruction errors for small momenta $p$ (compare $\Delta\rho_{[100]}$ in Fig.1 a) and b) and $\rho(k)$ shown in Fig. 2). The next change introduced into this method was the following. To make our results more consistent, instead of performing the reconstruction inside the same unit circle $p^{max}$ (for each plane $p_z$), we performed the reconstruction inside the unit sphere having a different unit circle ($p^{max}(p_z)=[(p^{max})^2-p_z^2]^{1/2}$) on each of the reconstruction planes. In order to shift densities in the whole space, we modified the expression for the density components $\rho_n(p_z,p)$ (these modifications will be described in detail in a further paper [12]) in the following:

$$\rho_n(p_z,p)=\Sigma_m a_n^m(2n+1+m)[p^{max}(p_z)/p^{max}]R_n^m(p/p^{max}(p_z))$$

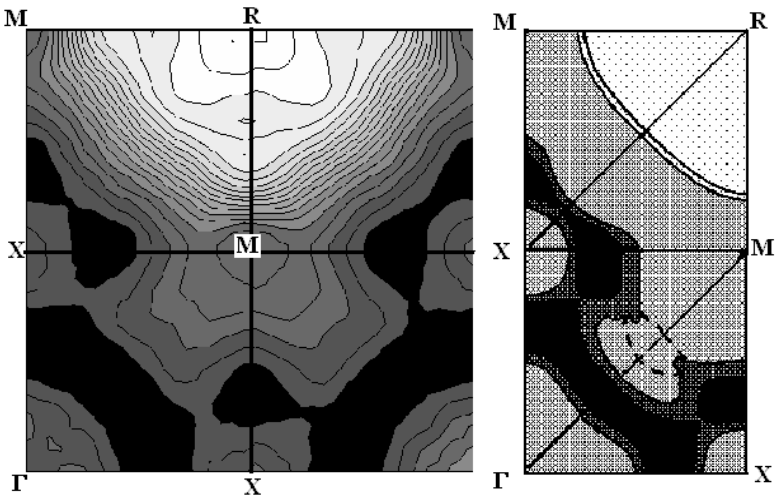

Fig. 2
Theoretical Fermi surface of TmGa$_3$ on the main symmetry planes [5] (right part) in the ferromagnetic configuration phase compared with $\rho(k)$ obtained by standard and modified CM (left and middle part, respectively).

It appears that our isodensities $\rho(k)$ (in the paramagnetic configuration phase) gather very close to the average FS of TmGa$_3$ in its ferromagnetic configuration phase. Moreover, due to the fact that the probability of positron annihilation depends strongly on the electron localisation, we observe the following: $1^0$ – There are more isodensities around the electron pocket centred at the R point in the 7th band than across the hole-like edges in the 6th band. This indicates that the electrons in the 6th band are more localised than those in the the 7th band; $2^0$ - the behaviour of the density in the extended $p$ space (see Fig. 3) show that the $4f^{12}$ electrons are strongly localised, i.e. they are below the bottom of the conduction band ($\rho(p)$ for higher momenta $p$ has very small values, i.e. the positron overlaps with the $4f$ electrons with a very small probability).

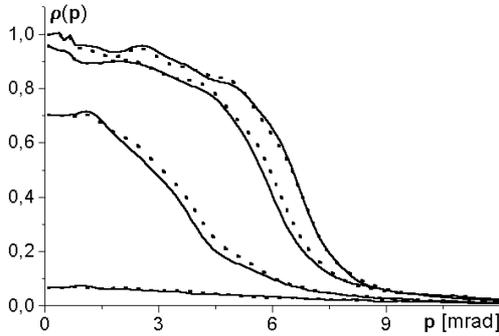

Fig. 3
Electron-positron densities in the extended momentum space along $\Gamma X$ (full line) and $\Gamma M$ (dotted line) on four reconstruction planes for $p_z=0$, $|G|/2$, $|G|$ and $3|G|/2$ (from the highest to the lowest curve, respectively). Here $G$ denotes reciprocal lattice vector [001].

Finally, by assuming that, although the e-p overlap integral depends on the degree of electron delocalisation, the FS discontinuities will remain where they were originally, we estimated the FS sizes (Table 1) by measuring the full width at half maximum (FWHM) of the appropriate high symmetry profiles.

Table 1. FS dimensions (in $[a.u.]^{-1}$) along high symmetry directions obtained from reconstructed densities, compared with theoretical band structure results [5]. Labels b7 and b6 denote the two conduction bands.

| Band | b7 | b7 | b7 | b6 | b6 | b6 | b6 | b6 |
|---|---|---|---|---|---|---|---|---|
| Direction | R-$\Gamma$ | R-X | R-M | $\Gamma$-M | $\Gamma$-X | $\Gamma$-R | X-M | X-$\Gamma$ |
| 2D ACAR | 0.291 | 0.289 | 0.306 | 0.144 | 0.17 | 0.128 | 0.08 | 0.1 |
| Theory [5] | 0.265 | 0.253 | 0.294 | 0.14 | 0.2 | No edge | 0.068 | 0.074 |

Our results show that the electron (7th band) and hole-like (6th band) FSs have a somewhat larger volume than predicted by the theory [5]. However, their qualitative shapes obtained from the experiment are very similar to the band structure results – see Fig. 2. Additionally, reconstructed e-p densities in the extended zone clearly show that the $4f^{12}$ states are strongly localised whereas folded densities $\rho(k)$ point out for different degrees of the electron localization in the 6th and 7th band.

**Acknowledgements**. We are grateful to the State Committee for Scientific Research (Republic of Poland, Grant No 2 P03B 083 16) for financial support.

For communication.(G.Kontrym-Sznajd) Tel.(48-71)3435021; Fax:(48-71)441029; mail:gsznajd@int.pan.wroc.pl